\documentclass[12pt]{article}
\usepackage{amsmath}
\usepackage{graphicx}
\begin{document}
\title{A debt behaviour model}
\author{Wenjun Zhang, John Holt}
\date{}
\maketitle{}

\begin{figure}[h!]
	\begin{center}
		\includegraphics[scale=0.5]{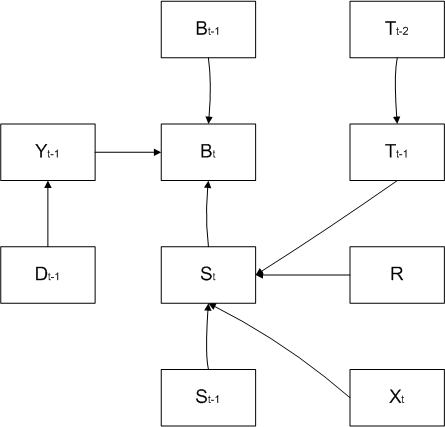}
		\end{center}
	\caption{This diagram depicts the underlying causal structure of the model.  See the text for the definitions of D,Y,B,T,S.}
	\label{fig:causal_structure}
\end{figure}

The model concerns the following random variables:
\begin{itemize}
\item A discrete Markov process $B_t$ which records the {\em behavioural state} of the debtor during the time period $t$ - measured in months.  The state is measured in the middle of each month.
\item A discrete-valued process $T_t$ which records the {\em strongest debt management intervention} that was applied to the debtor during the time period $t$.
\item $R$ an entity-specific variable, $R$ gives the final result of the debtor's most immediate previous debt case - NA, paid in full, liquidation/bankrupty, full write-off, partial write-off.
\item $X_t$ is the economic state at time period $t$.  This measure is obtained through clustering a pertinent collection of economic variables: change in CPI, change in unemployment, change in the average weekly wage, etc.  The underlying variables for $X_t$ are varying quarterly, so $X_t$ will be constant in blocks of three months.
\item $S_t$ is a latent discrete Markov process which categorizes debtors in a time period into the {\em behavioural scheme} that governs the generation of $B_t$.  The model supposes that $T_{t-1}$ influences $S_t$, and hence influences $B_t$ indirectly.
\item $D_t$ is a positive real-valued variable, given by $$D_t = \frac{\text{Debt amount at time $t$, including penalties and interest}}{\text{Largest amount of debt owed up to time $t$, excluding penalties and interest}}$$
\item $Y_t$ is a categorization of $D_t$ into $\{0,1\}$ - this is governed by a parameter $\alpha$ that needs to be inferred.  the notion is that as a debtor gets closer to being paid in full, its probability of making a large lump-sum payment to clear its debt may change.
\end{itemize}

We introduce a set of parameters as follows:
\begin{itemize}
\item $\alpha$: defined by $Y_t:=0$ if and only if $D_t\leq \alpha$.
\item $Q_S$: a list of transition matrices, one for each combination of values of $R,X_t,T_{t-1}$.
\item $\pi_S$: a list of initial probabilities, one for each combination of values of $R,X_t$.
\item $Q_B$: a list of transition matrices, one for each combination of values of $Y_{t-1}$ and $S_t$.
\item $\pi _B$: a list of initial probabilities, one for each value of $S_1$.
\end{itemize}

\begin{figure}[h]
	\begin{center}
		\includegraphics[scale=0.5]{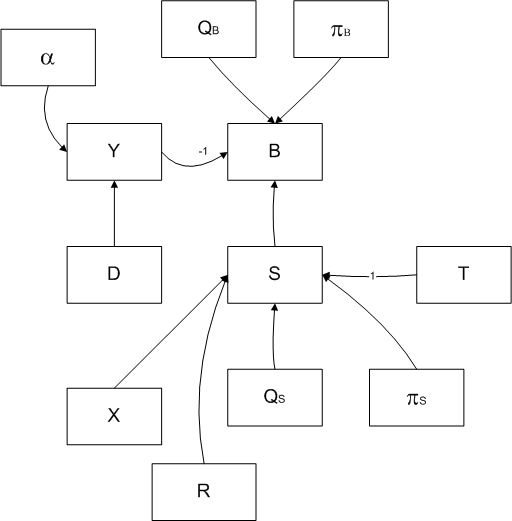}
		\end{center}
	\caption{This diagram depicts the underlying causal structure of the model, including the parameters. Refer to the text for definitions of the parameters $\pi_B,Q_B,\pi_S,Q_S, \alpha$}
	\label{fig:causal_structure_params}
\end{figure}

Figure \ref{fig:causal_structure_params} depicts the causal structure of the variables and the parameters - we have now expressed each of the variables as a vector of length as long as the number of observation periods.

Every debt case begins at a time period $u$ and ends at a time period $l$.  If the debt case is indexed by $i$, the the beginning is $u_i$ and the end is $l_i$.  There will be observations of $T_t$, $B_t$, $D_t$, and $X_t$ from $u_i$ through to $l_i$.

The log-likelihood of observing a single debt case is maximized when we maximize:
$$l_0 = \sum _{t=u+1}^{t=l}( \ln(Q_B^{Y_{t-1},S_t}(B_{t-1},B_t)) + \ln(Q_S^{X_t,R,T_{t-1}}(S_{t-1},S_t))) + \ln(\pi^{S_u}_B(B_u)) + \ln(\pi_S^{X_u,R}(S_u))$$

We apply the EM algorithm to $l_0$, taking the expected value of $l_0$ conditional on $\{B_t,X_t,D_t,T_t,R\}$ and the $k$-th iteration of the parameters $\{\alpha,Q_B,Q_S,\pi_B,\pi_S\}$, $\Theta^k$.

For this we define the {\em responsibilities} for each debt case, $i$, and time $t$, $t=u_i,\ldots,l_i$:

$$\gamma_{i,t}(s): = p(S_t=s | T_{u_i}^{l_i-1},X_{u_i}^{l_i},B_{u_i}^{l_i},R_i,D_{u_i}^{l_i-1})$$

for $t\geq u_i$; and for $t>u_i$,

$$\Gamma_{i,t}(p,q):=p(S_t=q, S_{t-1}=p | T_{u_i}^{l_i-1},B_{u_i}^{l_i},R_i, D_{u_i}^{l_i-1})$$

It is clear that $\gamma_{i,t}(s) = \sum _p\Gamma_{i,t}(p,s)$, or if $t=u_i$, $\gamma_{i,u_i}(s)=\sum_q \Gamma_{i,u_i+1}(s,q)$ - hence we need only compute $\Gamma_{i,t}$.

This is done using the Forward-Backward algorithm:

\section{Calculating $\Gamma_{i,t}$}

This calculation is standard, but we present it for completeness.

Define the following four sets of probabilities:
\begin{itemize}
\item $\pi_t(s) = p(S_t=s | T_{u}^{l-1},X_u^l,R,D_u^{l-1},B_u^l)$
\item $\pi _t'(s)=p(S_t=s | T_{u}^{t-1},X_u^t,R,D_u^{t-1},B_u^t)$, $t\geq u$.
\item $F_t(p,q) = p(S_{t-1}=p,S_t=q | T_{u}^{t-1},X_u^t,R,D_u^{t-1},B_u^t)$, $t>u$
\item $\Gamma_t(p,q) = p(S_{t-1}=p,S_t=q | T_{u}^{l-1},X_u^l,R,D_u^{l-1},B_u^l)$,$t>u$.
\end{itemize}

Then
\begin{eqnarray*}
F_t(p,q) &\propto & Q_B^{q,Y_{t-1}}(B_{t-1},B_t)Q_S^{T_{t-1},X_t,R}(p,q)\pi _{t-1}'\\
& = & (Q_B^{q,0}(B_{t-1},B_t)I_{[0,\alpha]}(D_{t-1})+ Q_B^{q,1}(B_{t-1},B_t)I_{(\alpha,\infty)}(D_{t-1}))Q_S^{T_{t-1},X_t,R}(p,q)
\end{eqnarray*}

and

$$\pi_t'(q) = \sum _p F_t(p,q)$$ with $\pi_u'(s)\propto \pi_B^s(B_u)\pi_S^{X_u,R}(s)$.  The normalizing constants can be found by noting that $\sum _{p,q}F_t(p,q)=1$ and $\sum _s \pi _u'(s)=1$.

Having obtained $F_t(p,q)$ (the {\em forward matrices}) we can calculate the {\em backward matrices} $\Gamma_t$ as follows:

Set $\Gamma_l=F_l$.

For $t<l$,
\begin{eqnarray*}
\Gamma _t(p,q) & = & p(S_{t-1}=p | S_t=q, T_{u}^{l-1},X_u^l,R,D_u^{l-1},B_u^l)p(S_t=q | T_{u}^{l-1},X_u^l,R,D_u^{l-1},B_u^l)\\
& = & p(S_{t-1}=p | S_t=q,T_{u}^{t-1},X_u^t,R,D_u^{t-1},B_u^t)\pi_t(q)\\
& = & F_t(p,q)\frac{\pi_t(q)}{\pi _t'(q)}
\end{eqnarray*}

\section{Update equations for the M-step}

The formulas that follow are the result of straightforward calculations.

\begin{eqnarray*}
Q_B^{s,y}(b,c) & = & \frac{\sum _i \sum _{t=u_i+1}^{l_i} \delta(B_{i,t}-c)\delta(B_{i,t-1}-b)\delta(Y_{i,t-1}-y)\gamma_{i,t}(s)}{\sum _i \sum _{t=u_i+1}^{l_i}\delta(B_{i,t-1}-b)\delta(Y_{i,t-1}-y)\gamma_{i,t}(s)}\\
\pi _B^s(b) & = & \frac{\sum _i \delta(B_{i,u_i}-b)\gamma_{i,u_i}(s)}{\sum _i \gamma_{i,u_i}(s)}\\
 Q_S^{T,R,X}(p,q) & = & \frac{\sum _i \sum _{t=u_i}^{l_i-1}\delta(T_{i,t}-T)\delta(R_i-R)\delta(X_t-X)\gamma_{i,t}(p)\gamma_{i,t+1}(q)}{\sum _i \sum _{t=u_i}^{l_i-1}\delta(T_{i,t}-T)\delta(X_t-X)\delta(R_i-R)\gamma_{i,t}(p)}\\
 \pi_S^{R,X}(s) & = & \frac{\sum _i \delta(R_i-R)\delta(X_{u_i}-X)\gamma_{i,u_i}(s)}{\sum_i \delta(R_i-R)\delta(X_{u_i}-X)}
\end{eqnarray*}

Note that $Q_B$ depends on an unknown value of $\alpha$.  The approach will be to fit $Q_B$ for a range of values of $\alpha$, and to choose the $\alpha$ that gives the maximum value to:

$$l_1= \sum _i \sum _{t=u_i+1}^{l_i}\sum_s\ln(Q_B^{s,0}(B_{i,t-1},B_{i,t})I_{[0,\alpha]}(D_{i,t-1}) + Q_B^{s,1}(B_{i,t-1},B_{i,t})I_{(\alpha,\infty)}(D_{i,t-1}))\gamma_{i,t}(s)$$

\end{document}